\documentclass{PoS}
\usepackage{wrapfig}
\usepackage{subcaption}
\usepackage{stackengine}

\title{Benchmarking COSI's detector effects engine}

\ShortTitle{Benchmarking COSI's detector effects engine}

\author{\speaker{Clio C. Sleator}, Steven E. Boggs, Jeng-Lun Chiu, Carolyn A. Kierans, Alex Lowell, John A. Tomsick, Andreas Zoglauer\\
        Space Sciences Laboratory, UC Berkeley\\
        E-mail: \email{sleator@berkeley.edu}}

\author{Mark Amman\\
	Lawrence Berkeley National Laboratory}

\author{Hsiang-Kuang Chang, Chao-Hsiung Tseng, Chien-Ying Yang\\
	Institute of Astronomy, National Tsing-Hua University, Taiwan}

\author{Chih-Hsun Lin\\
	Institute of Physics, Academia Sinica, Taiwan}

\author{Pierre Jean, Peter von Ballmoos\\
	IRAP Toulouse, France}

\abstract{The Compton Spectrometer and Imager (COSI) is a balloon-borne gamma-ray (0.2-5 MeV) telescope with inherent sensitivity to polarization. COSI's main goal is to study astrophysical sources such as $\gamma$-ray bursts, positron annihilation, Galactic nucleosynthesis, and compact objects. COSI employs a compact Compton telescope design utilizing 12 high-purity cross strip germanium detectors (size: $8\times8\times1.5$ cm$^3$, 2 mm strip pitch).\\
We require well-benchmarked simulations to simulate the full instrument response used for data analysis, to optimize our analysis algorithms, and to better understand our instrument and the in-flight performance. In order to achieve a reasonable agreement, we have built a comprehensive mass model of the instrument and developed a detailed detector effects engine, which takes into account the individual performance of each strip as well as the characteristics of the overall detector system. We performed detailed Monte-Carlo simulations with \emph{Cosima}/Geant4, applied the detector effects engine, and benchmarked the results with pre-flight calibrations using radioactive sources.  After applying the detector effects engine, the simulations closely resemble the measurements, and the standard calibration, event reconstruction, and imaging pipeline used for measurements can also be applied to the simulations.\\
In this manuscript, we will describe the detector effects engine, the benchmarking tests with calibrations, and the application to preliminary results from COSI's 46-day balloon flight in 2016.}

\FullConference{11th INTEGRAL Conference Gamma-Ray Astrophysics in Multi-Wavelength Perspective,\\
		10-14 October 2016\\
		Amsterdam, The Netherlands}

\begin{document}

\section{Introduction}
The Compton Spectrometer and Imager (COSI) is a balloon-borne, soft $\gamma$-ray (0.2-5 MeV) telescope designed to perform astrophysical observations. On May 17, 2016, COSI was launched from Wanaka, New Zealand on NASA's super pressure balloon and had a successful 46-day flight. The main science goals of the 2016 flight include measuring the polarization of extreme astrophysical environments such as $\gamma$-ray bursts, Galactic black holes, and active galactic nuclei, mapping the 511-keV positron annihilation line, and imaging diffuse emission from nuclear lines such as $^{26}$Al, $^{60}$Fe, and $^{44}$Ti. See \cite{carolynproc} for more details about the 2016 flight and the COSI science goals.

\begin{wrapfigure}{l}{0.45\textwidth}
	\centering
	\includegraphics[width=0.45\textwidth]{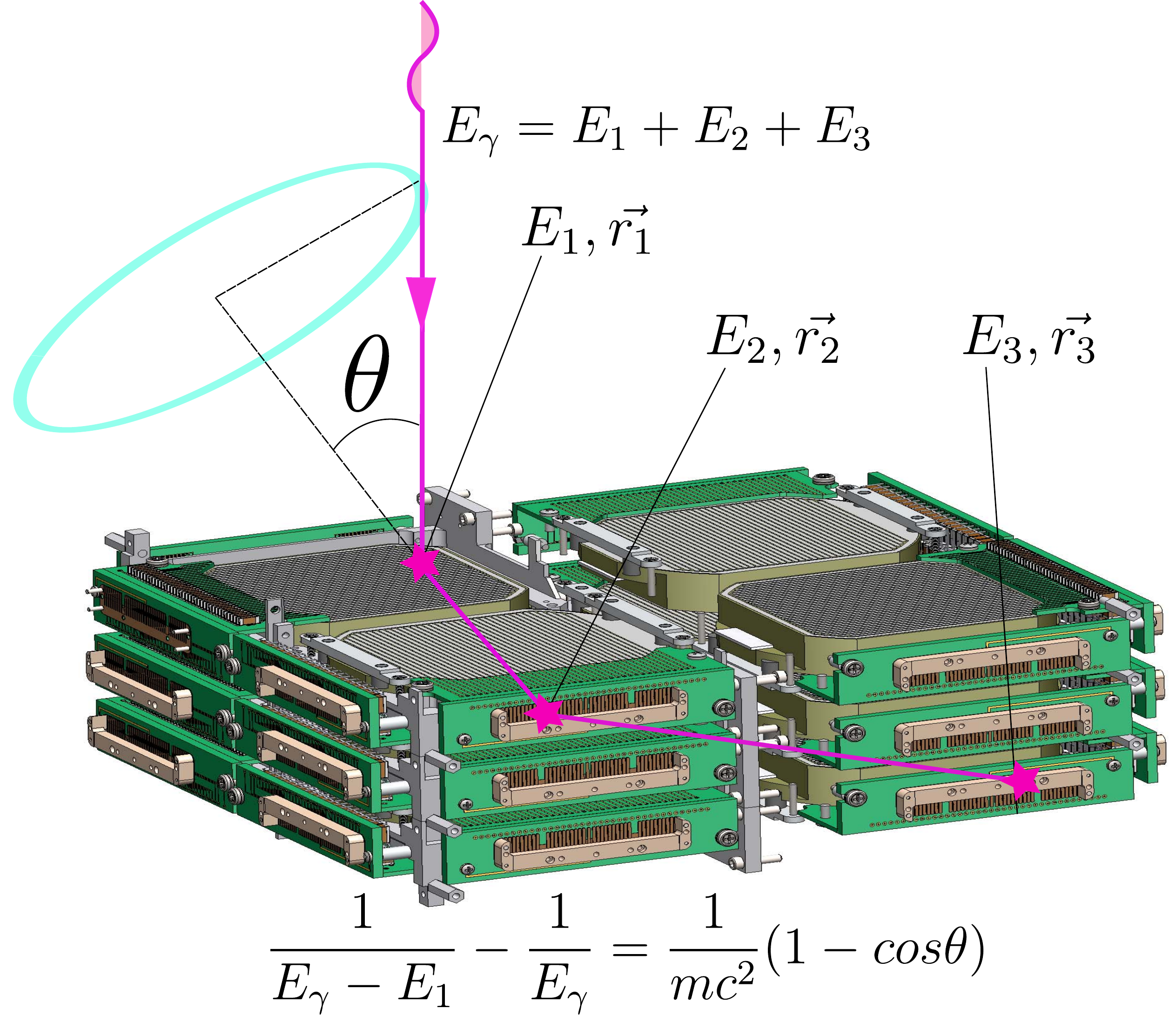}
\caption{An event with initial energy $E_{\gamma}$ undergoes two Compton scatters in the detectors at $r_1$ and $r_2$, depositing energies $E_1$ and $E_2$, and is then photobsorbed at $r_3$, depositing its remaining energy $E_3$. The first Compton scatter angle $\theta$ is calculated with the Compton equation as shown and the origin of the photon is somewhere on the blue event circle.\label{fig:compton}}
\end{wrapfigure}

COSI utilizes a compact Compton telescope design. Because Compton scattering is the dominating interaction process in the 0.2-10 MeV band for most detector materials, Compton telescopes are powerful tools for measuring soft $\gamma$-ray emission. Compact Compton telescopes have an active detector volume, in which a photon ideally undergoes multiple Compton scatters before being photoabsorbed (see Figure \ref{fig:compton}). The energy deposited at each interaction and the interaction position are used to determine the most probable interaction sequence using one of a variety of techniques \cite{boggs00} \cite{bayesian}. Once the interaction sequence is determined, the Compton equation and the detector response are used to constrain the origin of the photon to a ring on the sky.

\begin{wrapfigure}{R}{0.4\textwidth}
	\vspace{-20pt}
	\centering
	\includegraphics[width=0.4\textwidth]{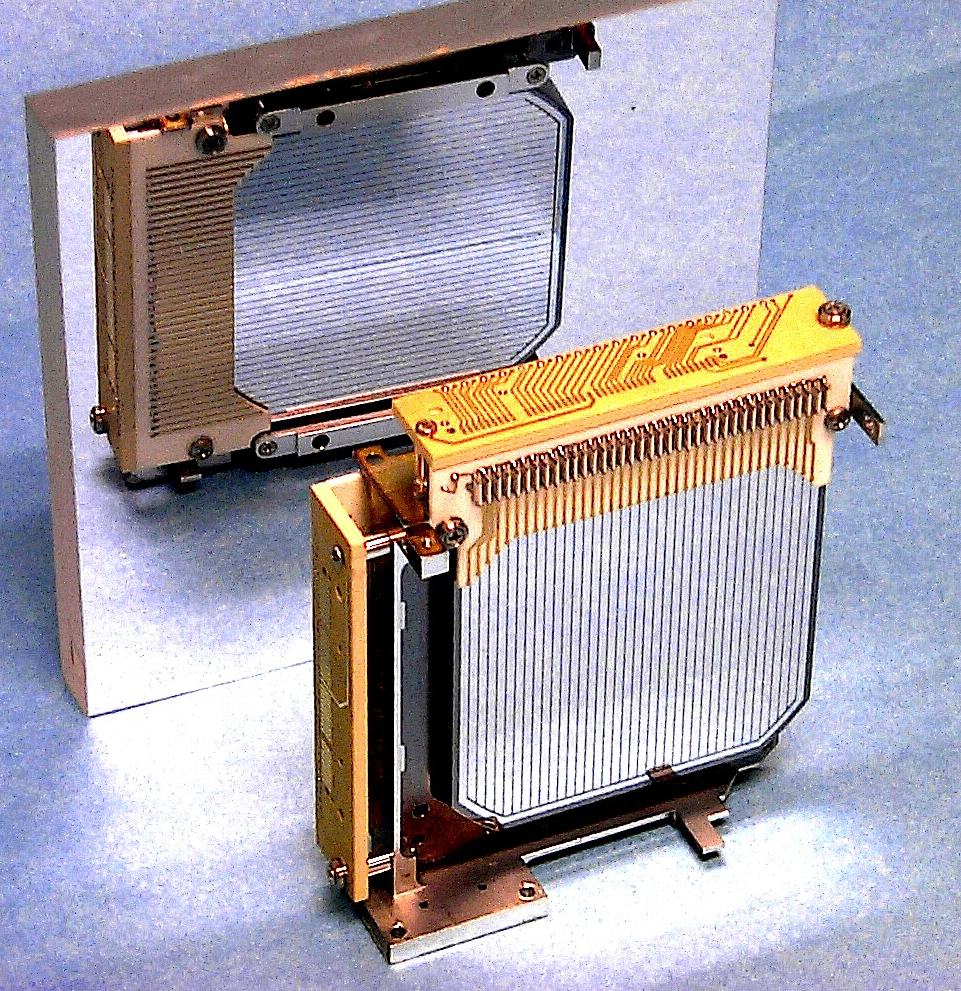}
\caption{A single GeD. The mirror behind the GeD shows the strips on the back which are laid out orthogonally to those on the front.\label{fig:GeDs}}
\vspace{-30pt}
\end{wrapfigure}

COSI's active detector volume is comprised of a $2\times2\times3$ array of high-purity cross strip germanium detectors (GeDs) \cite{amman}, each with a volume of $8\times8\times1.5$ cm$^3$ (see Figure \ref{fig:GeDs}). The anode and cathode electrodes on each side of each detector are segmented into 37 strips with a strip pitch of 2 mm. The strips on the anode are deposited orthogonally to those on the cathode so that the $x$ and $y$ position of the interaction can be determined using the positions of the triggered strips. A 2-mm guard ring surrounds the detector to prevent leakage current between the anode and the cathode. The guard ring also vetoes interactions that occur too close to the edge of the detector, where fringes in the electric field can degrade the detector response. The detectors are housed in a cryostat and are kept at cryogenic temperatures with a mechanical cryocooler, enabling ultra-long duration balloon flights. The cryostat bottom and sides are enclosed by a cesium iodide active anti-coincidence shield system.

Well-benchmarked simulations are fundamental for the COSI data analysis. Simulations enable us to understand the full instrument response, which is necessary for event reconstruction, imaging, and polarization analysis. we can also use simulations to benchmark and improve the data analysis pipeline by comparing the output of the analysis of simulated events with the idealized input. Additionally, simulations are necessary to better understand the calibration of the instrument and the instrument's in-flight performance.


As required for the simulations, we have built a comprehensive mass model of the instrument and developed a detailed detector effects engine (DEE) which applies the intrinsic detector performance to Monte Carlo simulations. The output of the DEE gives the best interaction energies and positions we can expect considering the detector energy and timing resolution, dead time, and other non-ideal effects. After being processed by the DEE, the simulated events go through the same calibration, event reconstruction, and image reconstruction as the real data. This manuscript describes the DEE and the preliminary benchmarking tests that we performed to ensure that our simulations closely match real data.



\section{Analysis Pipeline}

\begin{wrapfigure}{r}{0.4\textwidth}
\includegraphics[width=0.4\textwidth]{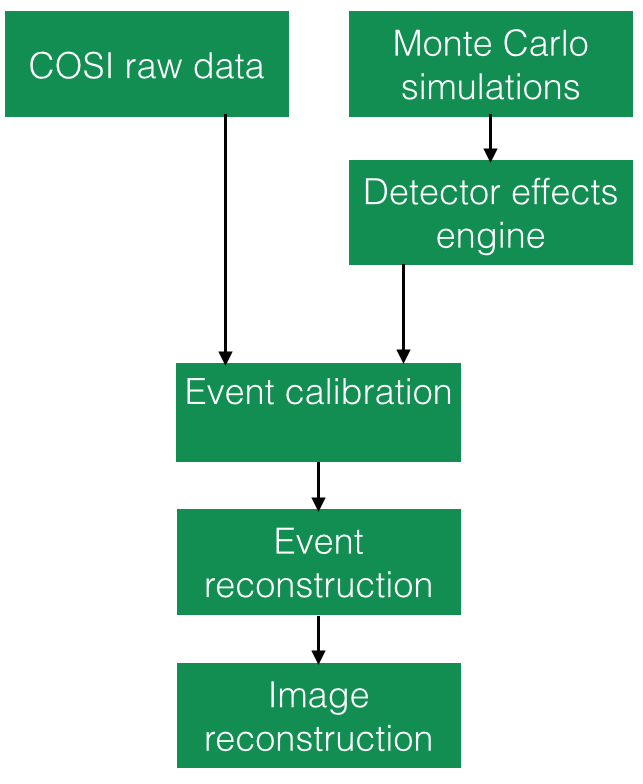}
\caption{A schematic of the COSI data analysis pipeline.\label{fig:pipeline}}
\vspace{-20pt}
\end{wrapfigure}

We use the Medium Energy Gamma-ray Astronomy library (MEGAlib) \cite{megalib}, software specifically designed to analyze data from Compton telescopes, for the COSI data analysis. Figure \ref{fig:pipeline} shows a schematic of the analysis pipeline. Both real data and simulations go through the event calibration, event reconstruction, and image reconstruction steps.

The event calibration converts the measured parameters of pulse height, pulse timing, strip ID, and detector ID into the physical parameters of energy and position, and consists of the following steps:
\begin{enumerate}
\item \emph{Energy calibration}: The pulse height is approximately proportional to the energy deposited in the interaction, so we can determine the pulse height to energy conversion using calibration sources with known line energies (e.g. $^{133}$Ba and $^{137}$Cs). We fit the pulse height-energy relation for each strip with an empirical model that accounts for any non-linear deviations. See \cite{kierans14} for more details.
\item \emph{Strip pairing}: We use the strip IDs to determine the $x$-$y$ interaction position in each detector. If there is only one interaction in a detector, then this process is straightforward as there is a signal on only one $x$ and one $y$ strip and the interaction position is where the strips meet. If, however, more than one interaction occurs, determining which $x$ and $y$ strips should be paired can be complicated. We pair the strips by comparing their energies, as the interaction should result in an equal amount of charge measured on the $x$ and $y$ strips.
\item \emph{Crosstalk correction}: Crosstalk is the influence of one electronics channel on another and causes an increase in measured energy on nearby strips. The effect is strongest on adjacent strips, but strips that are separated by one strip also exhibit this effect. Since the crosstalk effect is linear with energy, it is possible to correct for it, as described in \cite{markthesis}.
\item \emph{Depth calibration}: We determine the $z$ interaction position, or depth, by measuring the collection time difference (CTD), which is the difference between the collection times of the electrons on one side of the detector and the holes on the other side. See \cite{lowell16} for a discussion on calibrating the CTD-depth relation. For COSI, determining the depth is limited by the noise on the timing measurements of each strip.
\end{enumerate}

\noindent It is important to note that the DEE must invert all four of the event calibration steps, i.e. turn the physical parameters of energy and position into the measured parameters of pulse height, pulse timing, strip ID, and detector ID, so that the simulations resemble the measurements. This is described in Section 3.

After the event calibration, the event reconstruction determines the most probable interaction order, thus determining the initial photon direction to a ring on the sky \cite{boggs00} \cite{bayesian}. The image reconstruction uses iterative deconvolution techniques to go from individual photon rings to a source position on the sky \cite{mimrec}.

\begin{wrapfigure}{R}{0.5\textwidth}
\vspace{-25pt}
\includegraphics[width=0.5\textwidth]{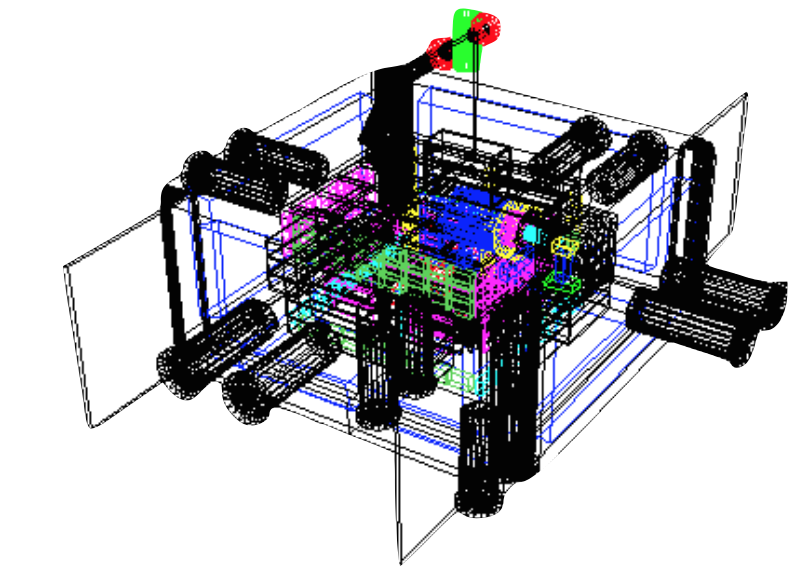}
\caption{The mass model of the detectors, cryostat, and shields shown with \emph{Geomega}. \label{fig:massmodel}}
\vspace{-10pt}
\end{wrapfigure}

\section{Simulation Pipeline}
As indicated in Figure \ref{fig:pipeline}, we perform Monte Carlo simulations that are processed by the DEE before going through the rest of the analysis pipeline. These steps are described in more detail in this section.

\subsection{Mass model and Monte Carlo simulations}
A realistic mass model of the detector geometry and nearby material is needed for simulations. The mass model dictates that the correct amount of material is at the correct position in the correct shape, and thus determines where in the detectors the simulated interactions occur. Because $\gamma$-rays can interact with passive material as well as with the active detectors and shields, it is important to model all objects near the detectors, including but not limited to the cryostat shell, cryocooler, and preamplifiers. A detailed description of the instrument materials modeled is presented in \cite{chiu}. The mass model is implemented in \emph{Geomega} \cite{cosima}, a MEGAlib program, and is shown in Figure \ref{fig:massmodel}. We use the mass model for our Monte Carlo simulations and all other analysis tools.

We perform Monte Carlo simulations with \emph{Cosima} \cite{cosima}, a $\gamma$-ray simulation tool in MEGAlib based on Geant4 \cite{geant4}. \emph{Cosima} performs Monte Carlo simulations of various source spectra and geometries and can perform simulations of space, balloon, and lab environments. As input, \emph{Cosima} requires the source position, which can be an astrophysical position or a position relative to the cryostat, and the source emission properties of energy spectrum, flux, and polarization. \emph{Cosima} outputs an event list describing interactions in the detectors as defined by the mass model.

\subsection{Detector effects engine}
The DEE begins with the idealized \emph{Cosima} output event list that describes the interactions in terms of their energy and position. Thus, the first step of the DEE is to invert the event calibration. Each interaction is decomposed into two \emph{strip hits} (one $x$, one $y$) representing a specific strip ID and its corresponding detector side, detector ID, pulse height, and timing.

\begin{figure}
	\centering
	\begin{minipage}{0.5\textwidth}
		\includegraphics[width=\textwidth]{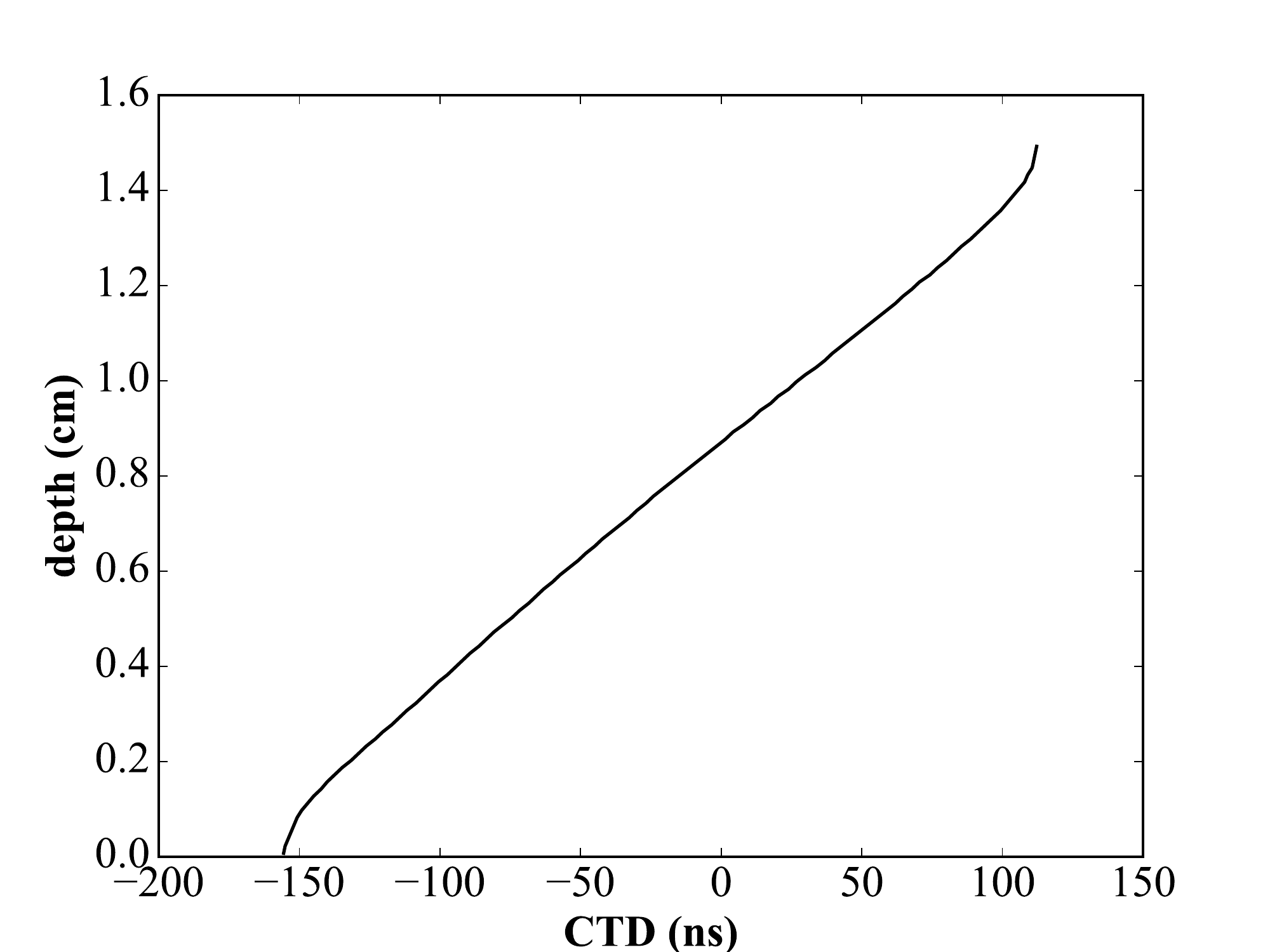}
		\caption{The relationship between CTD and depth for the pixel defined by $x$ strip 17 and $y$ strip 17 on detector 5.\label{fig:CTDdepth}}
	\end{minipage}
	\hfill
	\begin{minipage}{0.42\textwidth}
		\includegraphics[width=\textwidth]{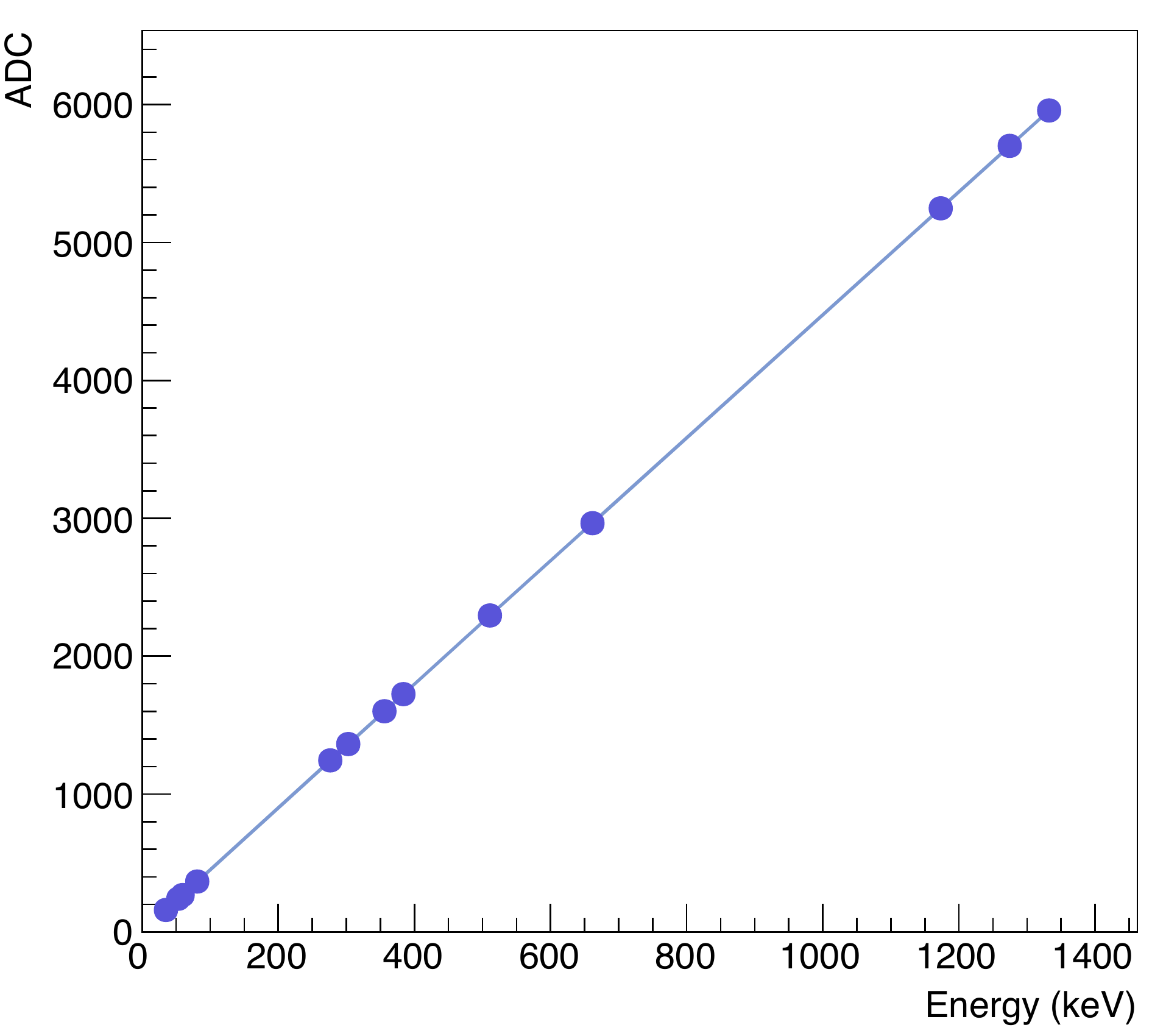}
		\caption{The relationship between pulse height (ADC) and energy of strip 7 on the positive ($x$) side of detector 7.\label{fig:ADCenergy}}
	\end{minipage}
\end{figure}

\emph{Position}$-$From the ($x,y,z$) position, we determine the detector ID, the $x$ strip ID, the $y$ strip ID, and the depth in the detector. We invert the depth calibration to convert depth to the collection time difference (CTD). Figure \ref{fig:CTDdepth} shows an example CTD-depth relation for one pixel \cite{lowell16}. Though only the CTD is used in the event calibration, each strip must be assigned an absolute timing such that the  output format of the DEE accurately mimics real data. We assign each strip an arbitrary timing while ensuring the correct CTD. We then apply Gaussian noise to the timing, using the Gaussian width measured in the depth calibration.

\emph{Energy}$-$We convert the energy into pulse height by inverting the energy calibration. Figure \ref{fig:ADCenergy} shows an example pulse height-energy relation for one strip \cite{kierans14}. To account for the detector energy resolution, we sample the pulse height from a Gaussian distribution with the mean equal to the true energy and the width as measured in the energy calibration.

\emph{Shield and guard ring vetoes}$-$If an interaction occurs in the shields and is temporally coincident with an event in the detectors, the event is discarded. A shield and detector event are considered coincident if the shield event occurs between 0.7 and 1.1 $\mu$s after the detector event triggers, as set in the read-out electronics. Similarly, an event in a specific detector is discarded if a guard ring interaction in the same detector occurs between 3.4 and 4 $\mu$s after the inital event triggers.

\begin{figure}
\begin{subfigure}[]{0.51\textwidth}
	\centering
	\includegraphics[width=\textwidth]{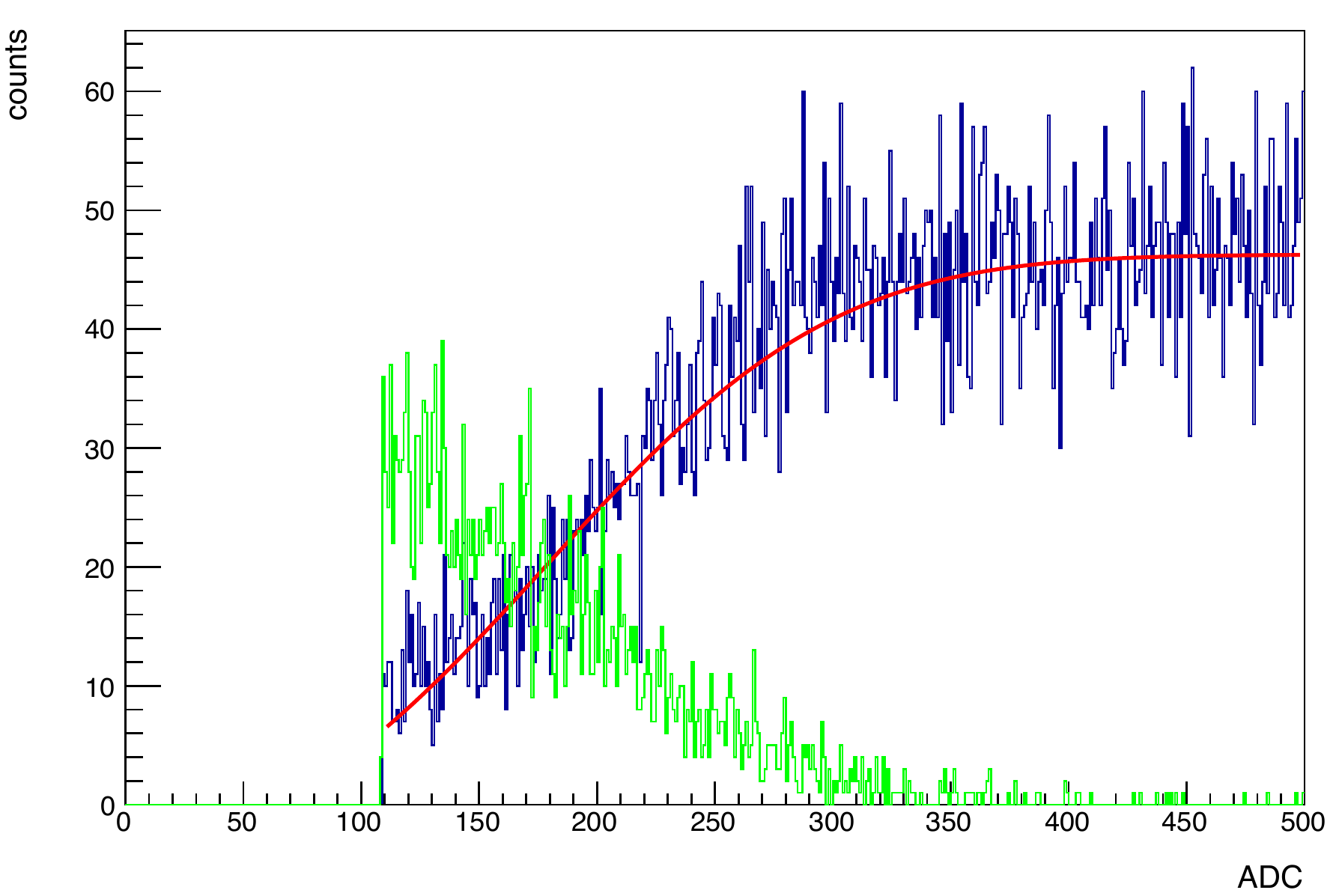}
	\caption{}
\end{subfigure}
\begin{subfigure}[]{0.51\textwidth}
	\centering
	\includegraphics[width=\textwidth]{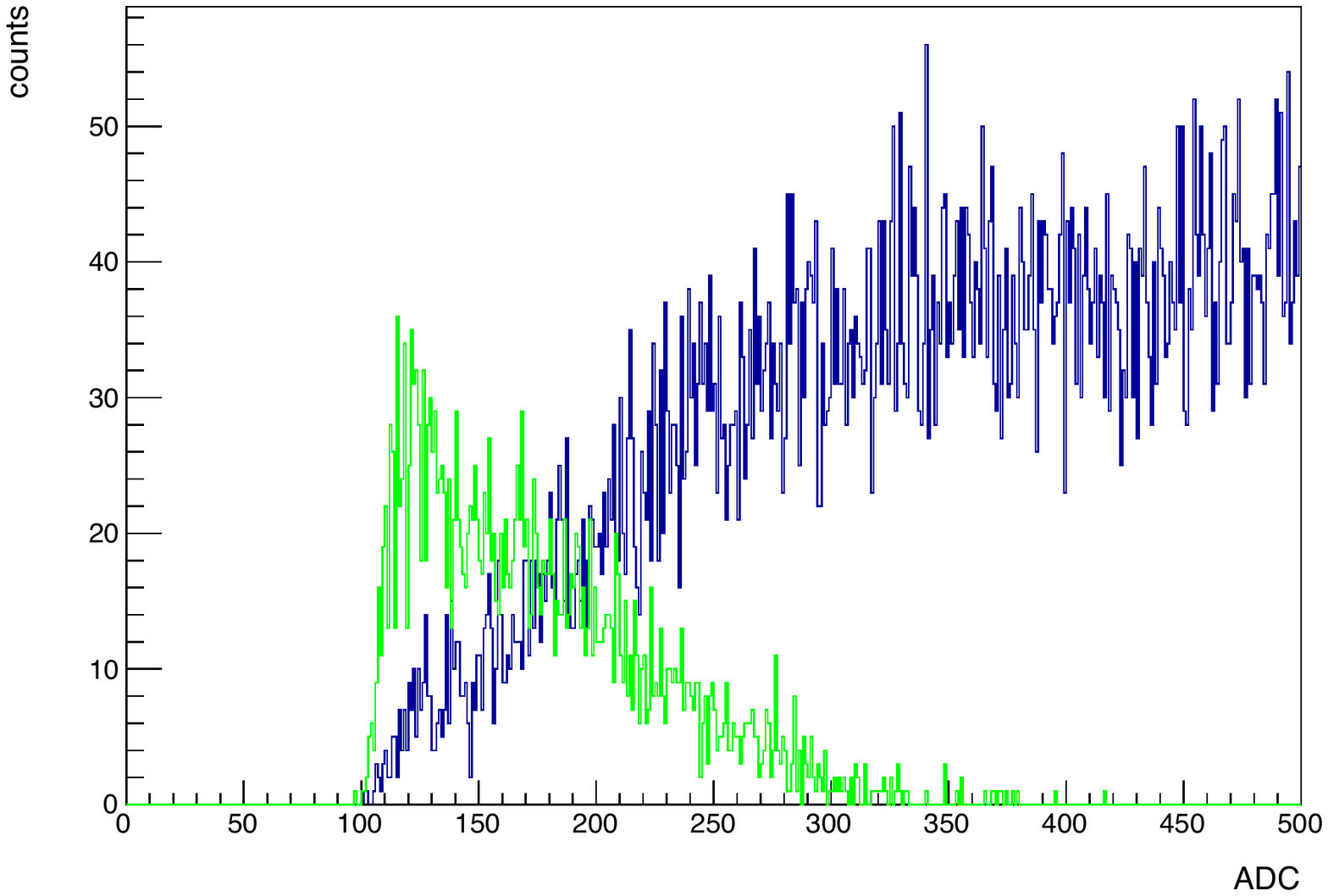}
	\caption{}
\end{subfigure}
\caption{The low-energy portion of the spectrum of strip 5 on the negative ($y$) side of detector 0. (a) Shows the calibration spectrum and (b) the simulated spectrum. The green spectrum is made up of energy-only events while the blue spectrum is made up of events with energy and timing. The cutoff at ADC (pulse height) = 110 ($\sim 20$ keV) denotes the energy threshold. The red line in (a) shows the fit to the error function.\label{fig:thresholds}}
\end{figure}

\emph{Thresholds}$-$Each strip has a timing threshold of $\sim40$ keV, below which timing information is not triggered, and an energy threshold of $\sim20$ keV, below which nothing is triggered. We calibrated the energy and timing thresholds of each strip individually by considering two separate spectra: one of energy-only events, and one of energy-and-timing events (see Figure \ref{fig:thresholds}). Because the energy channel has low noise, both spectra have a sharp cut-off at the energy threshold. The timing channel is noisy, so we modeled the low-energy regime of the energy-and-timing spectrum with an error function. The error function is the integral of a Gaussian with a mean equal to the timing threshold and a width describing the noise. To apply the thresholds to the simulated strip hits, we discard any strip hits with energies below the energy threshold.  If the strip hit energy is above the energy threshold but below the timing threshold $\pm$ Gaussian noise, we remove the timing information so that the strip hit becomes an energy-only event. Additionally, a small fraction ($\sim1\%$) of the strips are dead; any simulated strip hits that occur on these strips are removed.

\emph{Trigger conditions}$-$A real event is only processed and saved if there is at least one strip on each side of the detector that has both energy and timing information. Once dead strips and thresholds are taken into account, not all simulated events meet this criteria; those that do not are discarded.

\emph{Shield and detector dead time}$-$The shield dead time is paralyzable and thus fairly straightforward to model: if the shields are dead when another interaction occurs, the dead time is extended. By measuring the shield count rate and percent live time, we determined that the shield dead time per event is $\sim 1.7$ $\mu$s. The detector dead time is much more complicated and is primarily due to the read out electronics. The detector dead time is non-paralyzable, meaning that if a new event occurs while the detector is dead, that event is neither triggered nor does it extend the dead time. As a first order model, we calculate the dead time per event by taking the inverse of the maximum count rate of each detector. We determined the maximum count rate by measuring the count rate when all of the calibration sources were above the cryostat at once, saturating the detectors. This count rate was about $\sim 1700$ cps for each detector, resulting in a dead time per event of $\sim 600$ $\mu$s. After an event in a detector, the detector is dead for 600 $\mu$s; if another event occurs in the same detector before the 600 $\mu$s pass, that event is discarded.

\emph{Crosstalk}$-$We are in the process of building crosstalk into the DEE. Crosstalk is the influence of nearby strips on each other and causes the measured energy to increase. To simulate crosstalk, we invert the calibrated correction as described in Section 2. 

\emph{Charge sharing and charge loss}$-$When interactions occur in the gap between two strips, the DEE currently assigns the interaction to the closest strip. In the real detectors, however, it is possible for charge to be collected on both strips, a phenomenon referred to as charge sharing. When charge sharing occurs, some charge can be lost in the gap between the strips. Additionally, one of the strips can collect an amount of charge below the energy threshold. Both of these types of charge loss lead to a tail on the low energy side of the spectral peak. We are working on adding these effects to the DEE. 

Because they affect the energy of the strip hits, crosstalk and charge loss have an effect on strip pairing (described in Section 2). If the energy of corresponding $x$ and $y$ strip hits do not match well, it is more difficult for strip pairing to produce correct results. This consequence to strip pairing is another reason to add crosstalk and charge loss to the DEE.

\begin{figure}
\begin{subfigure}[]{0.5\textwidth}
	\centering
	\includegraphics[width=\textwidth]{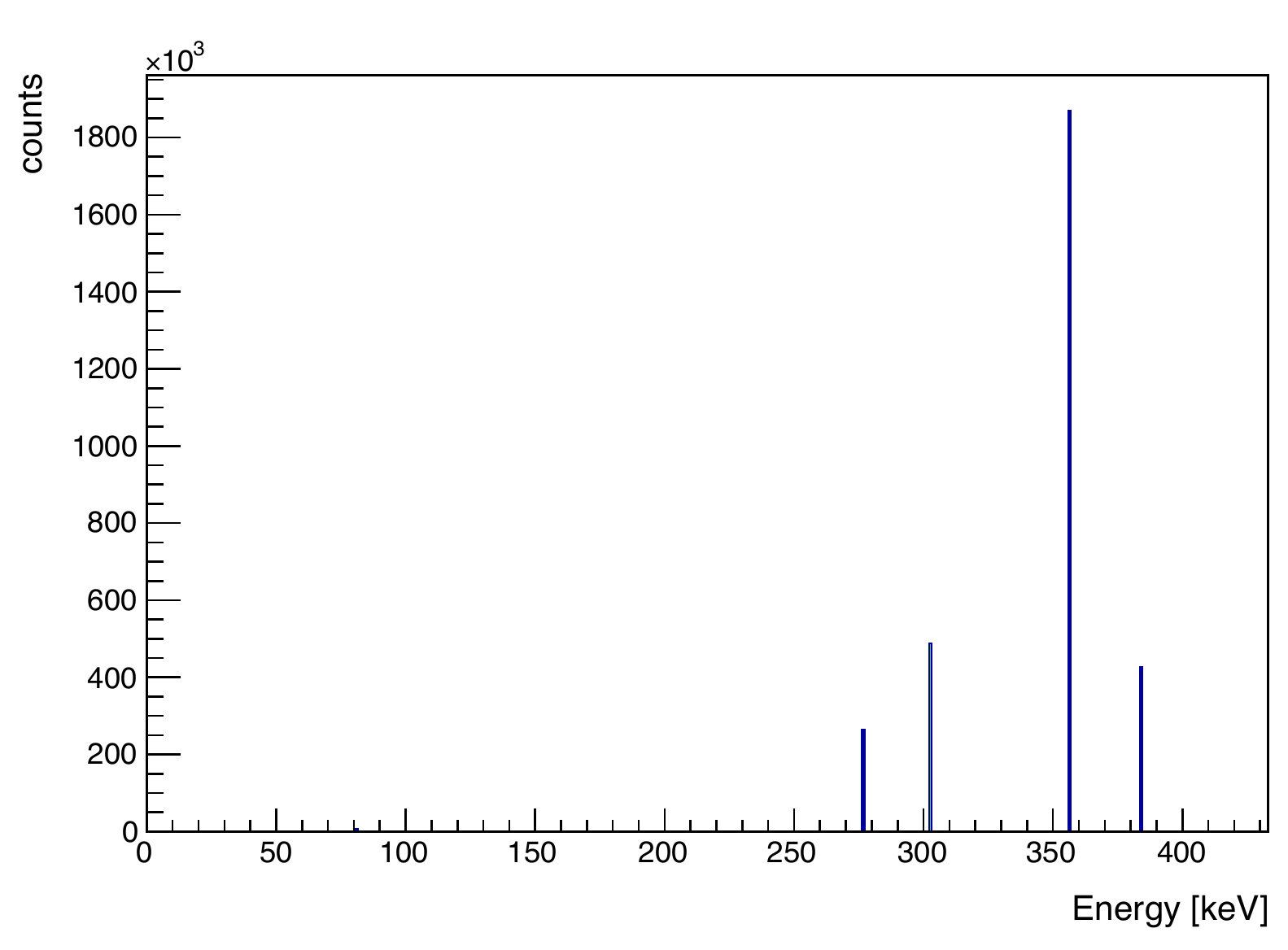}
	\caption{}
\end{subfigure}
\begin{subfigure}[]{0.5\textwidth}
	\centering
	\includegraphics[width=\textwidth]{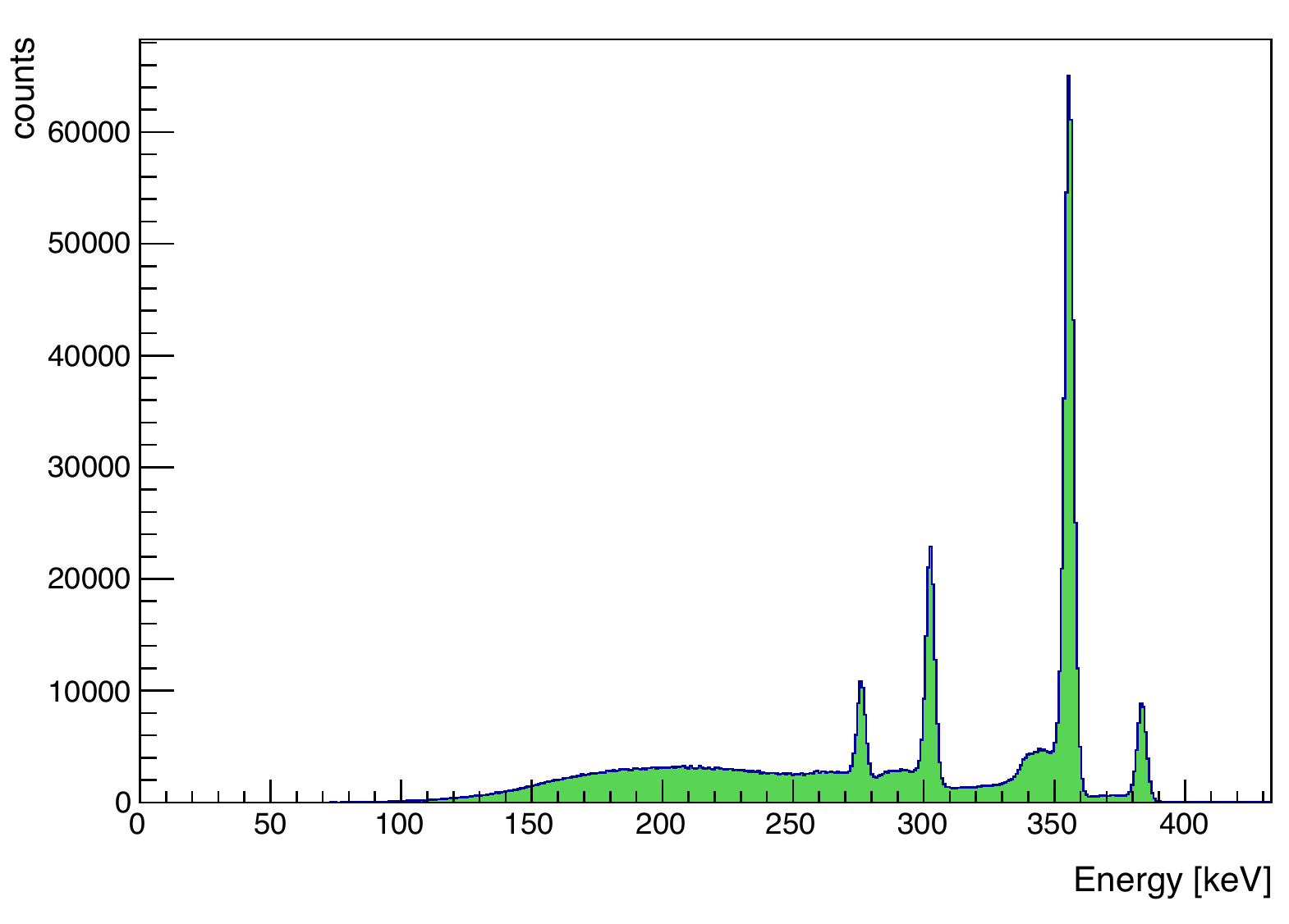}
	\caption{}
\end{subfigure}
\caption{(a) Spectrum of the idealized \emph{Cosima} output of a $^{133}$Ba source. (b) Spectrum of the same simulation after being processed by the DEE. Note the difference in scale.\label{fig:ideal}}
\end{figure}

\emph{Effects of the DEE}$-$Figure \ref{fig:ideal} shows a simulated $^{133}$Ba spectrum with and without the DEE. The most evident consequences of the DEE to the spectrum are the changes in peak height, the addition of finite energy resolution, and the presence of the continuum due to incompletely absorbed events.

\begin{figure}
\begin{subfigure}[]{0.5\textwidth}
	\centering
	\stackinset{l}{0.42in}{t}{0.18in}{\includegraphics[width=0.38\textwidth]{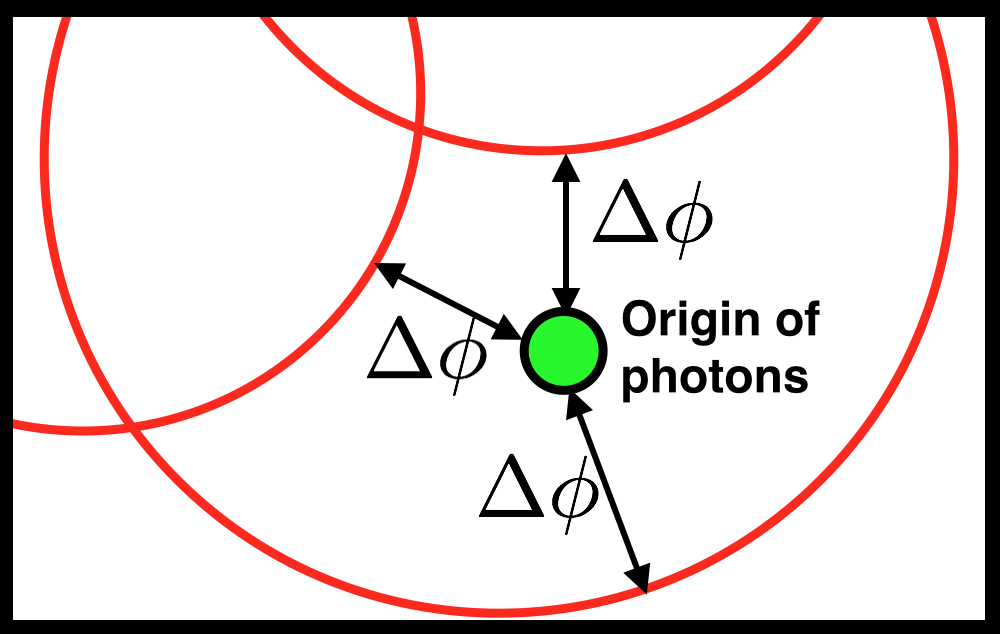}}{\includegraphics[width=\textwidth]{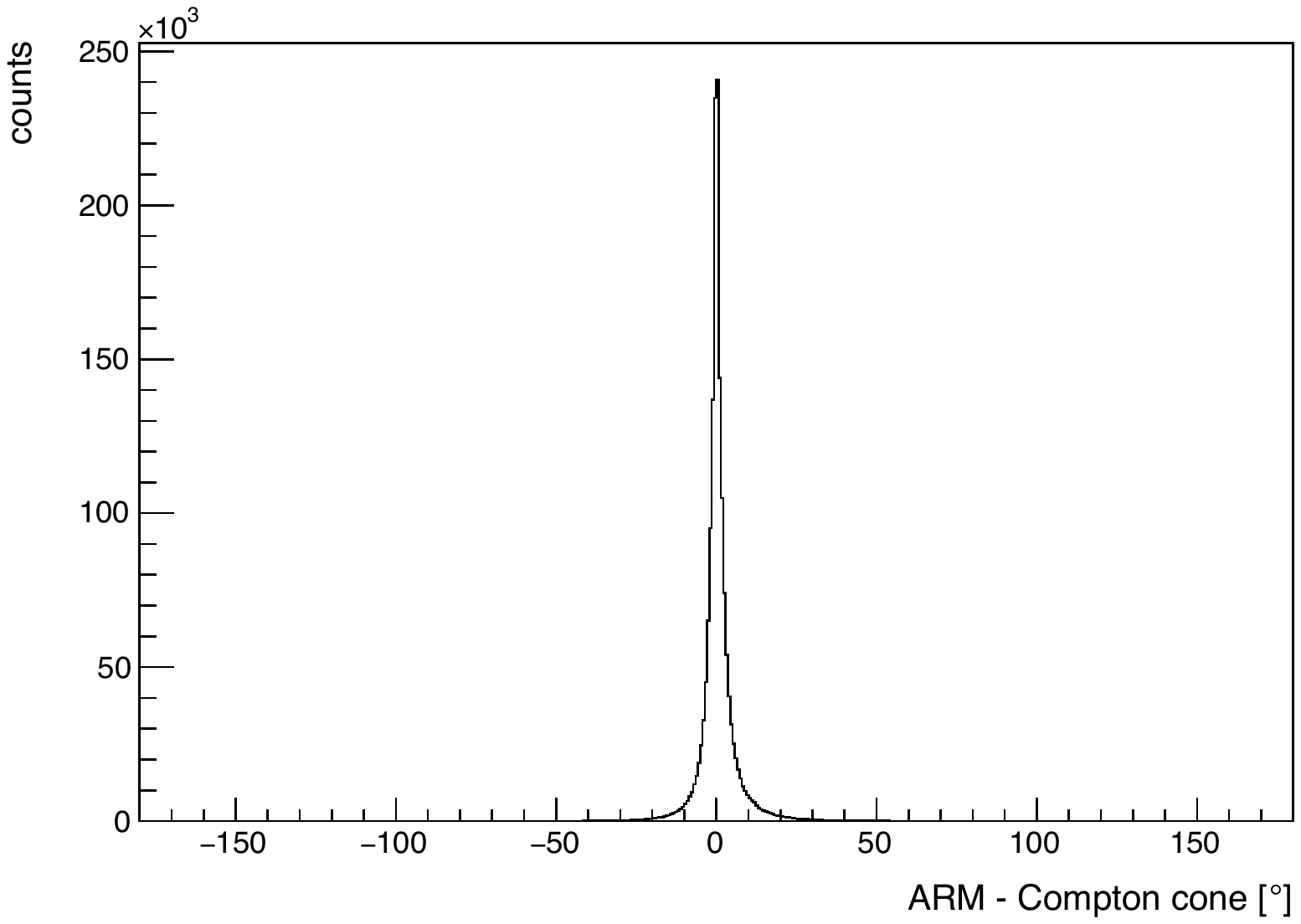}}
	\caption{\label{fig:ideal_arm}}
\end{subfigure}
\begin{subfigure}[]{0.5\textwidth}
	\centering
	\includegraphics[width=\textwidth]{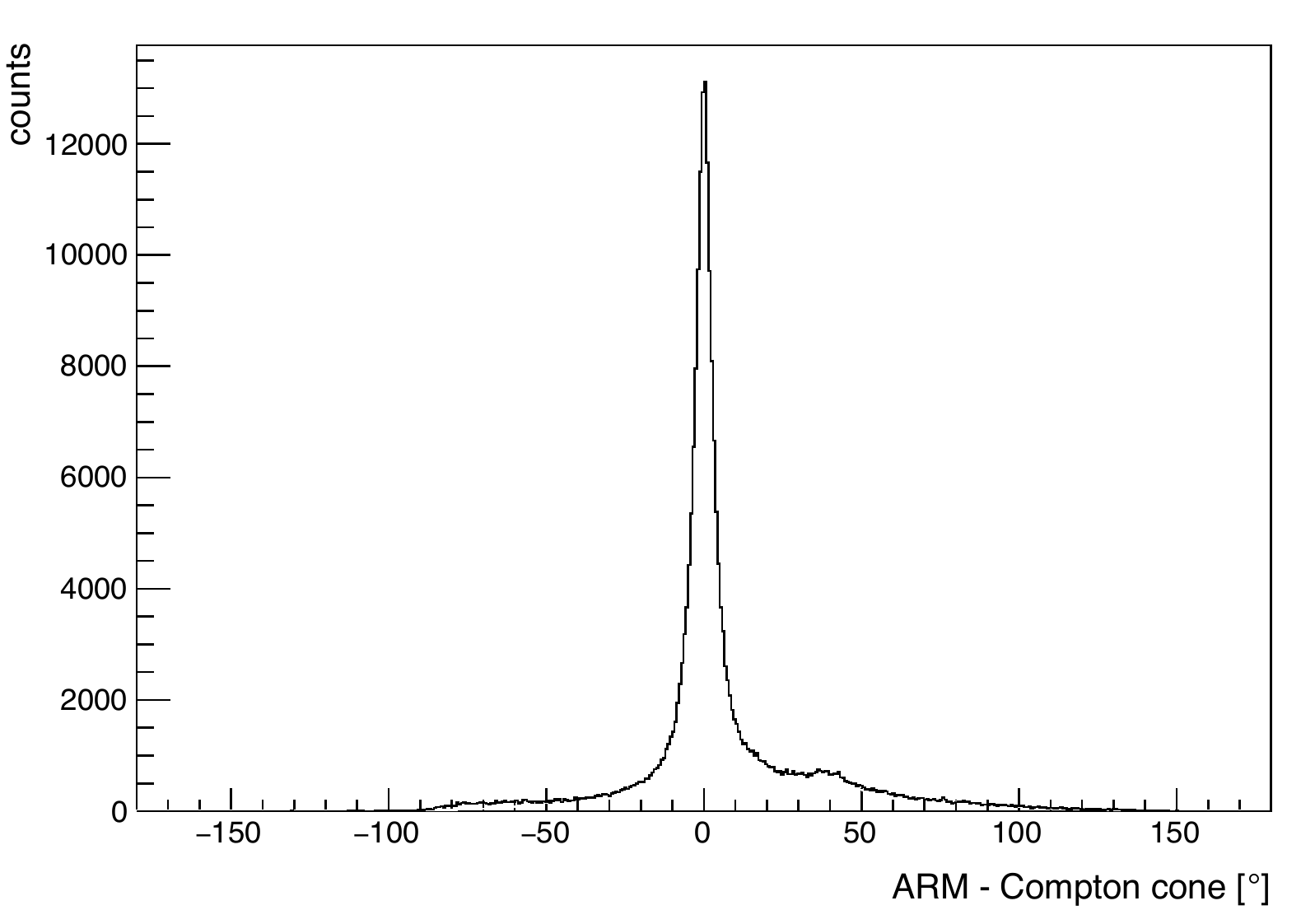}
	\caption{\label{fig:deearm}}
\end{subfigure}
\caption{(a) ARM of the idealized \emph{Cosima} output of a $^{133}$Ba source. (b) ARM of the same simulation after being processed by the DEE. Note the difference in scale. The inset in (a) demonstrates how the ARM is calculated: the green circle is the source position and each red arc represents a section of a single photon's Compton cone. The plataeu at $\sim50^{\circ}$ in (b) is due to incorrectly reconstructed events.\label{fig:idealarm}}
\end{figure}

Another useful tool for benchmarking is the angular resolution measure (ARM), which is used to characterize the angular resolution of a Compton telescope. The ARM is the distribution of the smallest angular distance between the known origin of the photon and each Compton cone (see the inset in Figure \ref{fig:ideal_arm}). The FWHM of the ARM defines the instrument angular resolution. Figure \ref{fig:idealarm} shows a simulated ARM for the $^{133}$Ba line at 356 keV with and without the DEE. The ARM is sensitive to energy and position resolution as well as the results of the event reconstruction; the effect of incorporating the finite position and energy resolution into the DEE is shown by the width of the distribution in Figure \ref{fig:deearm}. The width of the idealized ARM in Figure \ref{fig:ideal_arm} is solely from Doppler broadening \cite{bayesian}.


\section{Preliminary Simulation Benchmarking}
To benchmark the DEE, we compare the simulations to calibration data taken in the lab. For our calibration data, we use radioactive sources with known energies and activities and place them in specific locations relative to the cryostat. These sources can be accurately simulated with \emph{Cosima} and are then processed by the DEE, followed by the event calibration and event reconstruction as shown in Figure \ref{fig:pipeline}. We compare the spectra, angular resolution, and other aspects of the simulations to the calibration data. In this section, we show preliminary comparisons between data and simulations of two calibration sources ($^{133}$Ba and $^{137}$Cs), each centered above the cryostat, to illustrate the status of the DEE.




\begin{figure}
\begin{subfigure}[]{0.5\textwidth}
	\centering
	\stackinset{l}{0.45in}{t}{0.25in}{\includegraphics[width=0.5\textwidth]{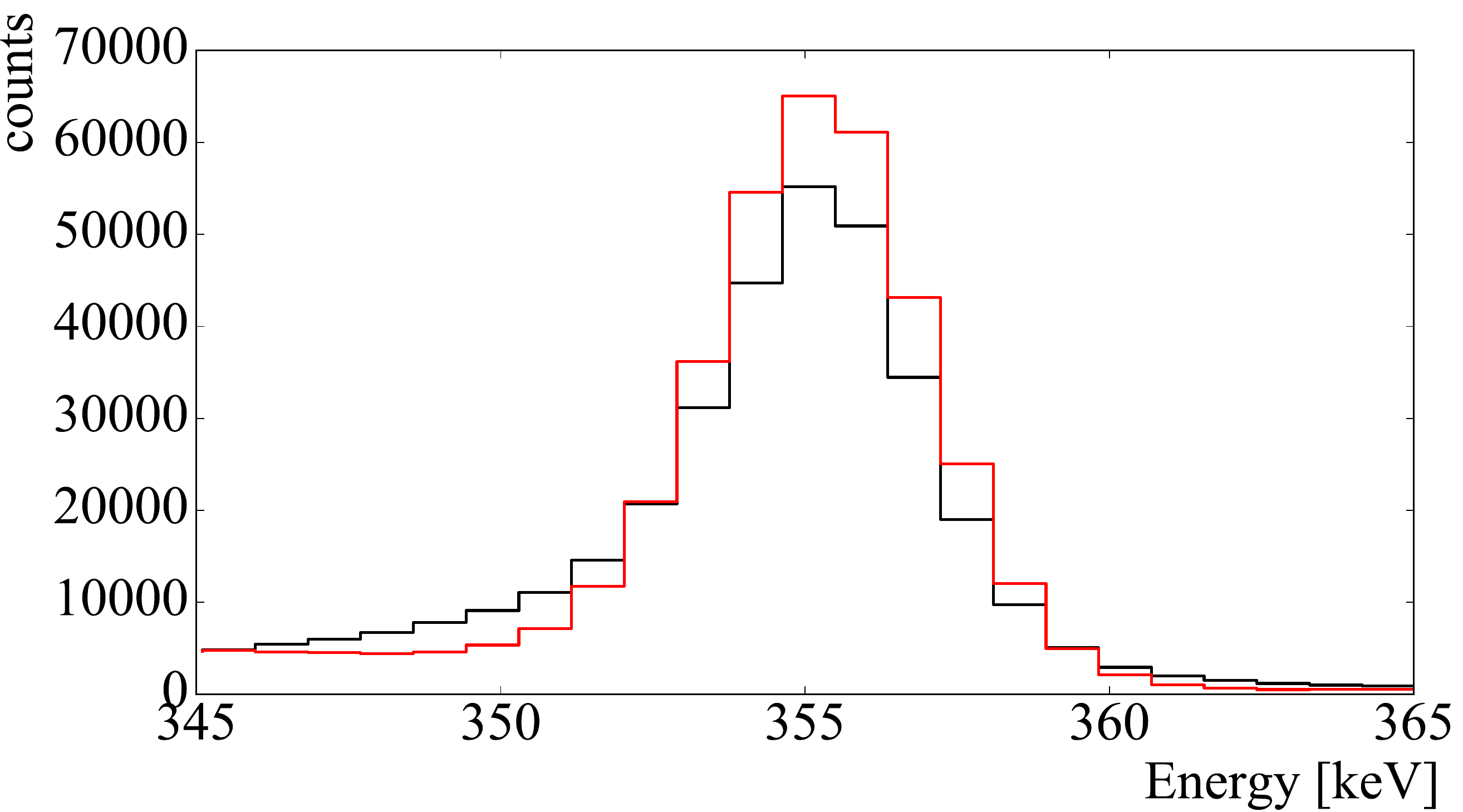}}{\includegraphics[width=1.05\textwidth]{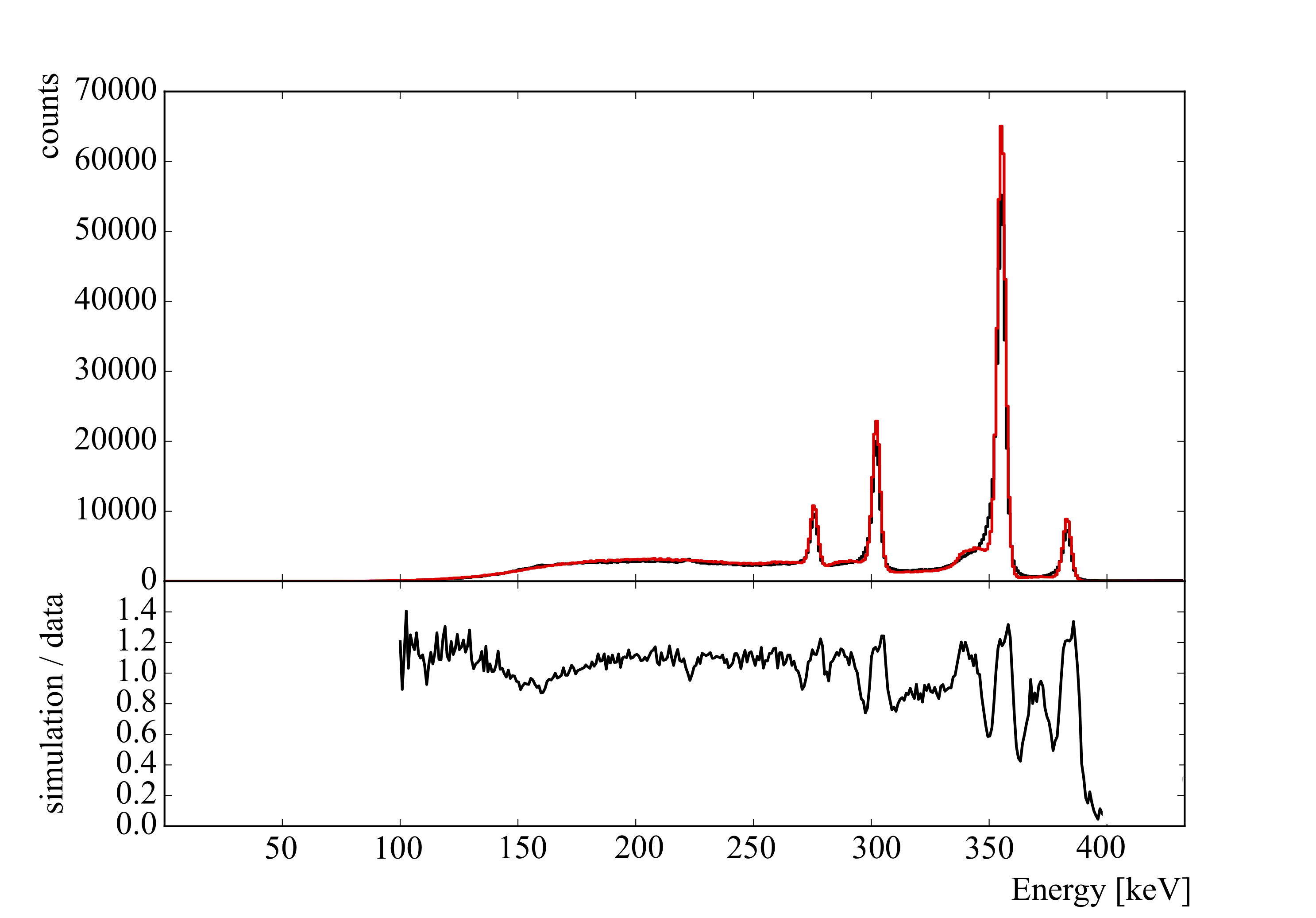}}
	\caption{}
\end{subfigure}
\begin{subfigure}[]{0.5\textwidth}
	\centering
	\stackinset{l}{0.45in}{t}{0.25in}{\includegraphics[width=0.6\textwidth]{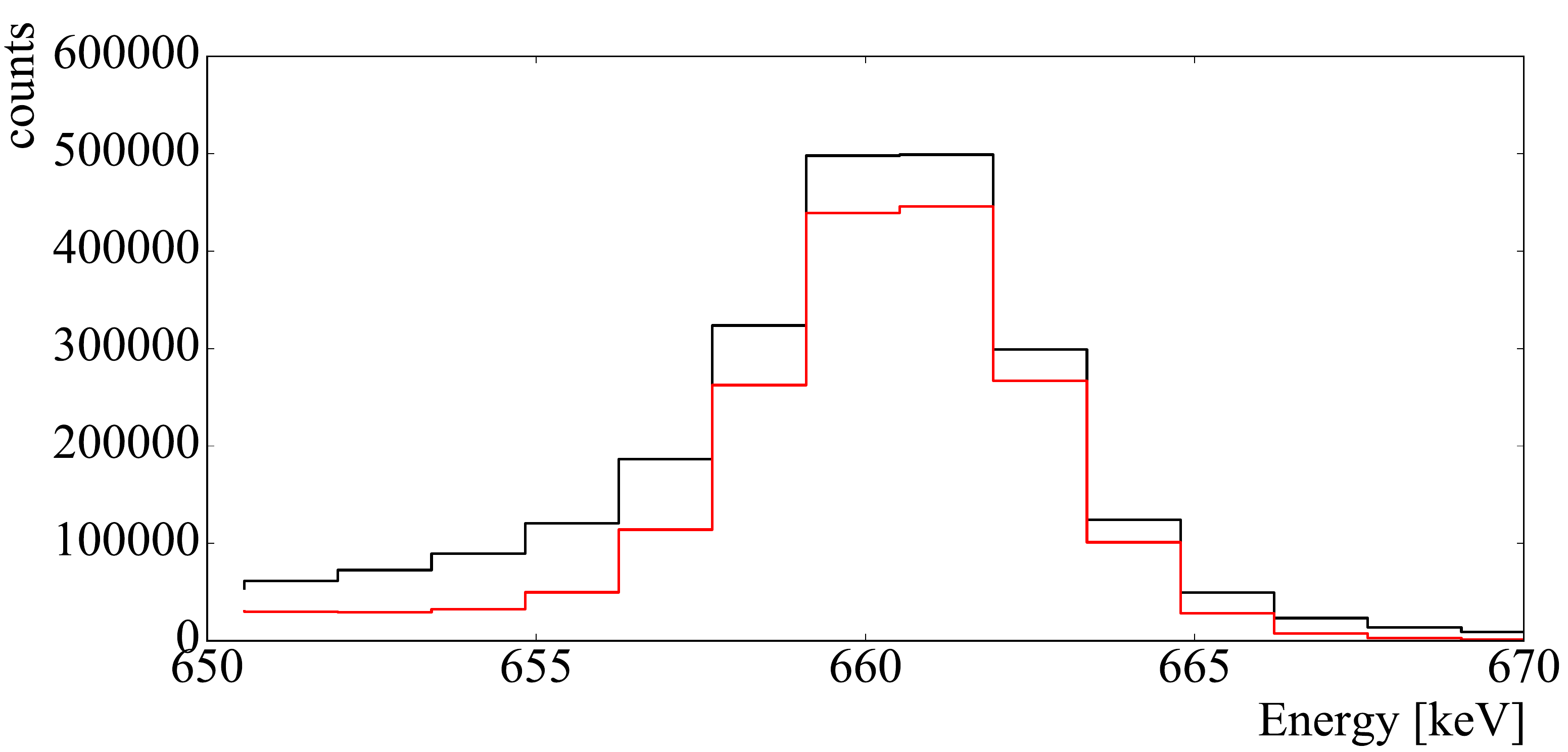}}{\includegraphics[width=1.05\textwidth]{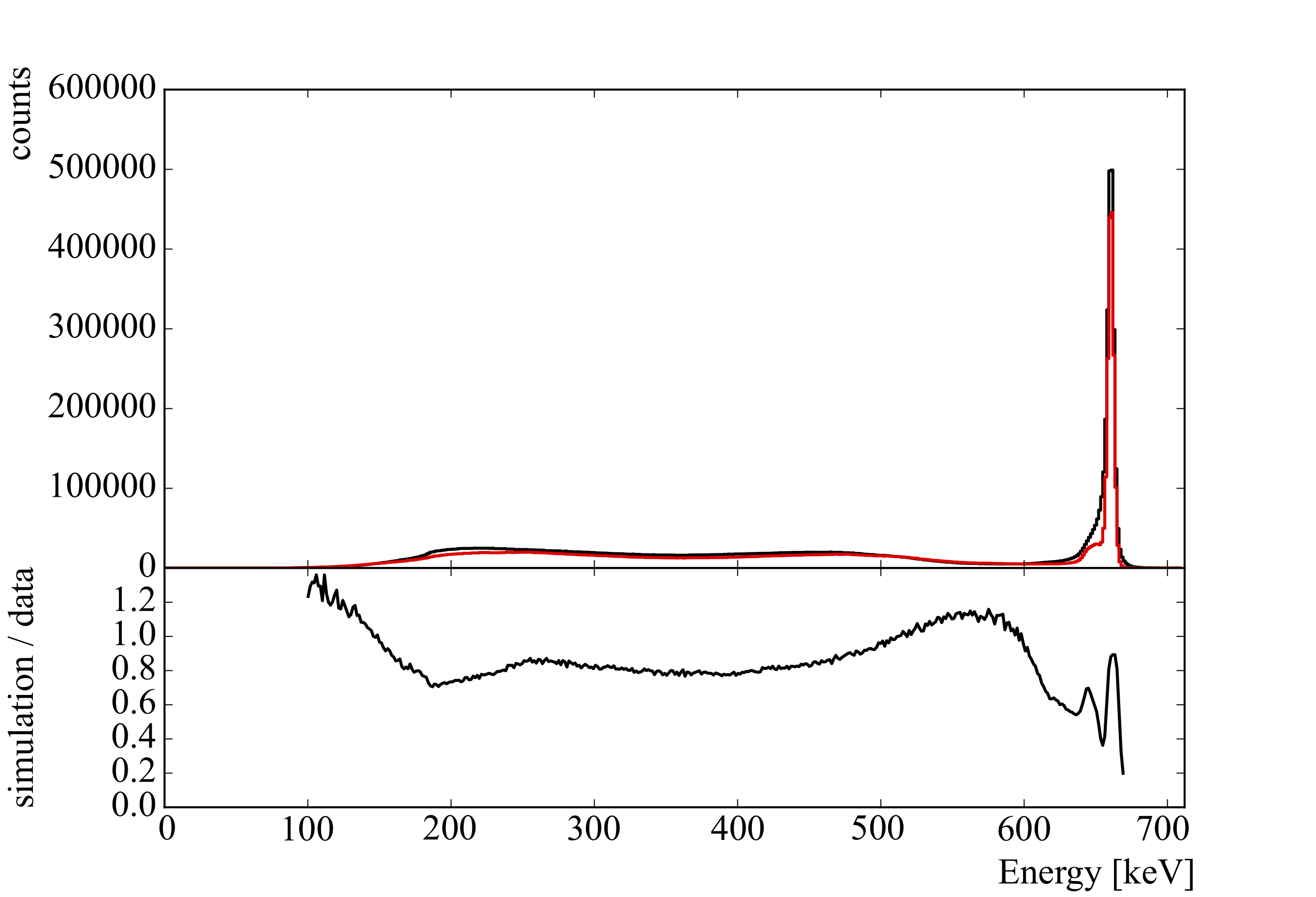}}
	\caption{}
\end{subfigure}
\caption{(a) The comparison between a simulated (red) and real (black) $^{133}$Ba spectrum. (b) The comparison between a simulated and real $^{137}$Cs spectrum. The inset shows a close-up image of the largest peak (at 356 keV for $^{133}$Ba and 662 keV for $^{137}$Cs). The lower panel shows the ratio of simulation counts to calibration data counts for each bin. \label{fig:spectrum}}
\end{figure}

Figure \ref{fig:spectrum} shows a comparison of the spectra of both sources. While the overall peak shape is well matched, there is some discrepancy in the tail below the peak and in the number of counts within the peak. The difference in the tail is most likely due to crosstalk and charge loss effects that we have not yet taken into account in the DEE. Work to include these effects is ongoing. We are also investigating the small difference between the number of counts within the calibration and simulation peaks.

\begin{figure}
\begin{subfigure}[]{0.5\textwidth}
	\centering
	\includegraphics[width=\textwidth]{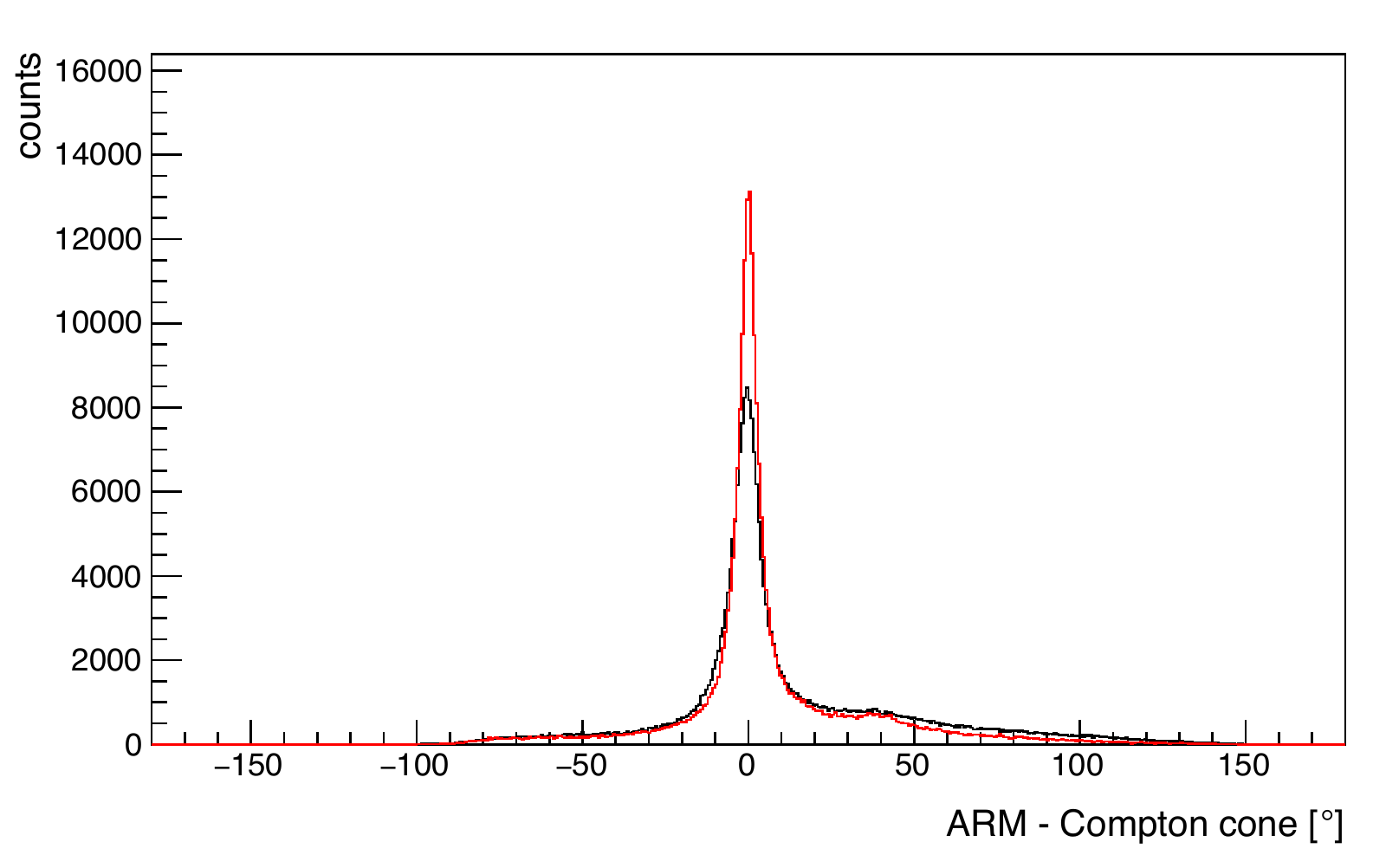}
	\caption{\label{fig:armBa}}
\end{subfigure}
\begin{subfigure}[]{0.5\textwidth}
	\centering
	\includegraphics[width=\textwidth]{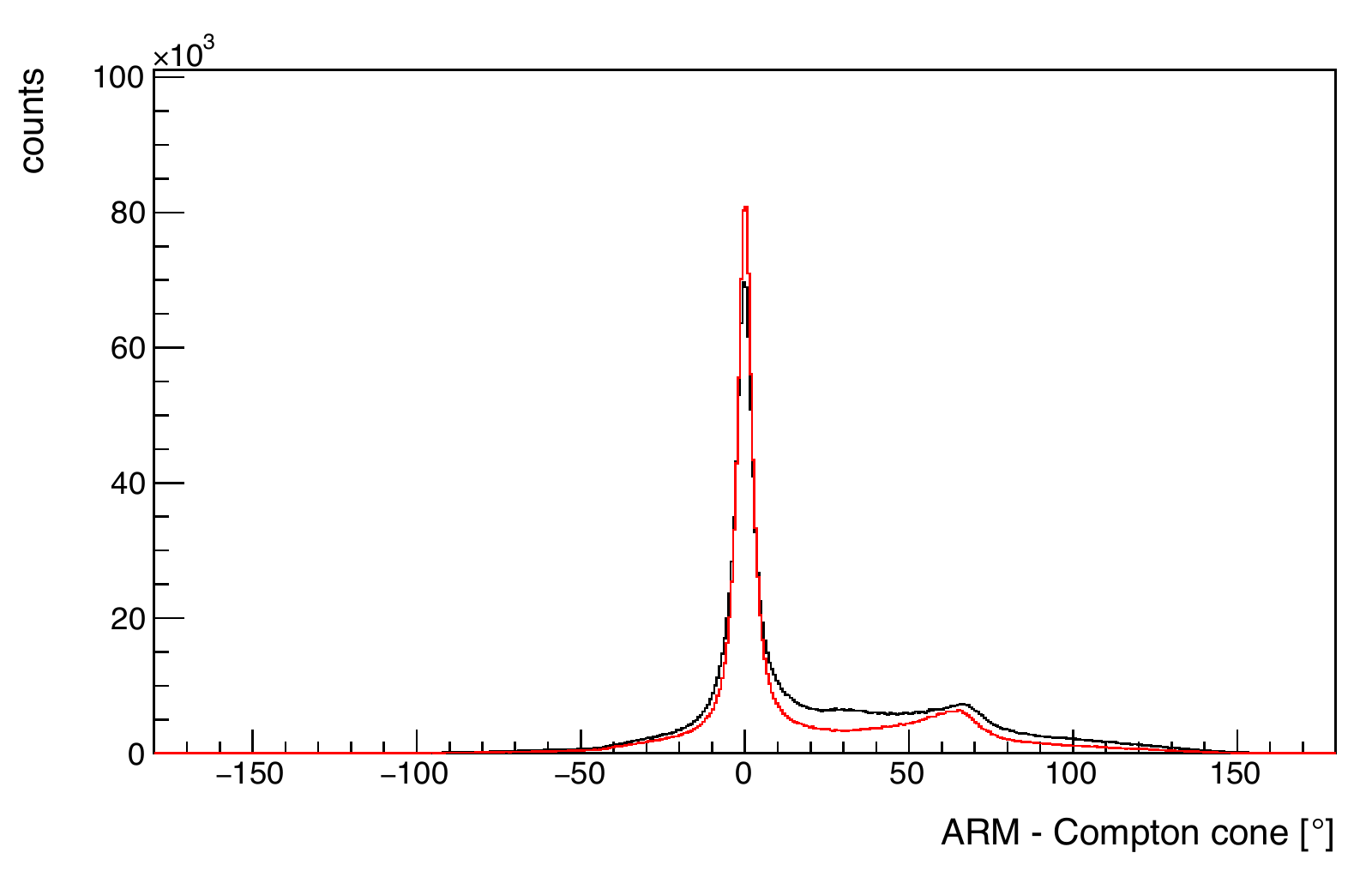}
	\caption{\label{fig:armCs}}
\end{subfigure}
\caption{(a) The comparison between a simulated (red) and real (black) $^{133}$Ba ARM. (b) The comparison between a simulated and real $^{137}$Cs ARM. The plateau at $\sim50^{\circ}$ is due to incorrectly reconstructed events.\label{fig:armdistribution}}
\end{figure}

Figure \ref{fig:armdistribution} shows a comparison between the simulated and real ARM at 356 keV$-$the strongest line in the $^{133}$Ba spectrum (Figure \ref{fig:armBa})$-$and at 662 keV, the only line in the $^{137}$Cs spectrum (Figure \ref{fig:armCs}). To compute the ARM at a specific energy, we select events that have energies within $E_{\rm peak}\pm1.5 \sigma$. There is reasonable but not perfect agreement between the real (FWHM $=6.0^{\circ}$) and simulated (FWHM $=5.2^{\circ}$) angular resolution at 662 keV. There is a larger discrepancy, however, between the real (FWHM $=9.5^{\circ}$) and simulated (FWHM $=6.8^{\circ}$) angular resolution at 356 keV. As the ARM is sensitive to the energy and position resolution and event reconstruction, all steps of the analysis pipeline can affect the ARM. Adding crosstalk and charge loss, currently missing from the DEE, will likely worsen the simulated angular resolution. It is also likely that improvements to the event calibration, which will affect the event reconstruction and ARM, can better the measured angular resolution.



\section{Conclusions}

We have described the status of the DEE and how we use calibration data taken in the lab to benchmark our simulations. We have made significant progress towards developing an accurate model of the instrument. Even so, there are still some discrepancies between our simulations and the calibration data, and work is ongoing to address these. Adding known detector effects such as crosstalk and charge loss in addition to more accurately modeling the detector dead time will most likely improve the agreement between simulations and calibrations. Any further discrepancies could be indicative of problems in the event calibration step of the analysis pipeline. 

With an accurate DEE, we will be able to optimize the data analysis pipeline and correctly simulate the detector response to sources seen during the 2016 flight, both necessary steps towards rigorously analyzing the COSI flight data.

\section*{Acknowledgements}

Support for COSI is provided by NASA grant NNX14AC81G.

\end{document}